# Plasma turbulence measured by fast sweep reflectometry on TORE SUPRA

F. Clairet[&], L. Vermare[&], S. Heuraux[€], G. Leclert[#]

[&]Association Euratom-CEA sur la fusion, DSM/DRFC/SCCP C.E. Cadarache, 13108 Saint-Paul-lès-Durance, France
[€]LPMIA, Université Henri Poincaré Nancy I, BP 239, 54506 Vandoeuvre-lès-Nancy, France
[#] LPIIM, CNRS-Univiversité de Provence, Faculté de St Jérôme, case 321, 13397 Marseille, France

Electronic mail : frederic.clairet@cea.fr

**Abstract**

Traditionally devoted to electron density profile measurement we show that fast frequency sweeping reflectometry technique can bring valuable and innovative measurements onto plasma turbulence. While fast frequency sweeping technique is traditionally devoted to electron density radial profile measurements we show in this paper how we can handle the fluctuations of the reflected signal to recover plasma density fluctuation measurements with a high spatial and temporal resolution. Large size turbulence related to magneto-hydrodynamic (MHD) activity and the associated magnetic islands can be detected. The radial profile of the micro-turbulence, which is responsible for plasma anomalous transport processes, is experimentally determined through the fluctuation of the reflected phase signal.

## I. Introduction

Plasma turbulence is among the most important issues, which have to be addressed to appreciate the plasma equilibrium questions of magnetic fusion experiments. Problems of plasma discharge sustainment [1] can be destabilized by MHD activity, which can drive damaging fast electron towards the inner wall and/or conduct major plasma disruption. For example, plasma confinement regimes with full wave current drive experiment, specially used to achieve long duration plasma discharges, are particularly exposed to MHD stability problems. Moreover, anomalous radial transport is enhanced by instabilities and the major mechanisms and/or the dominant driving forces for these turbulent processes are never clearly identified. Due to such theoretical difficulties there are issues in which experiments could help to clarify these studies so that innovative and comprehensible diagnostic measurements are required. Reflectometry is a diagnostic permitting detailed research on this particular topic and benefits on a well-established technique.

## II. Fast frequency sweeping reflectometry

Tore Supra is equipped with fast frequency sweeping (20 μs) X-mode heterodyne reflectometers [2] dedicated to electron density profile measurements and covering the frequency range from 50 to 110 GHz. These reflectometers provide electron density profile measurements routinely and automatically on a shot to shot basis. Fast and repetitive sweeps (with 5 μs dead time between sweeps), named burst mode technique, provide an equivalent sampling of 40 kHz at a given probing frequency and allow fast plasma events to be measured. The heterodyne and sine/cosine detection allows separate measurements of phase and amplitude of the reflected signal with a S/N sensitivity of about 40 dB. Reflectometry relies on the fact that, as an electromagnetic wave propagates through the plasma, its phase is shifted due to the departure of the local refractive index from the vacuum value. At a certain critical density corresponding to the cut-off layer, this refractive index goes to zero and then the probing wave is reflected. For the reflectometers of TORE SUPRA, the wavenumber ($k$) of the probing waves are ranging from $k_0$ (10 to 23 cm$^{-1}$) into the vacuum to zero when the cut-off layer is reached. In addition to the reflection at the cut-off, backscattering due to the density fluctuations and ranging up to $2k_0$ (according to the Bragg's law) occurs all along the propagation path (figure 1).

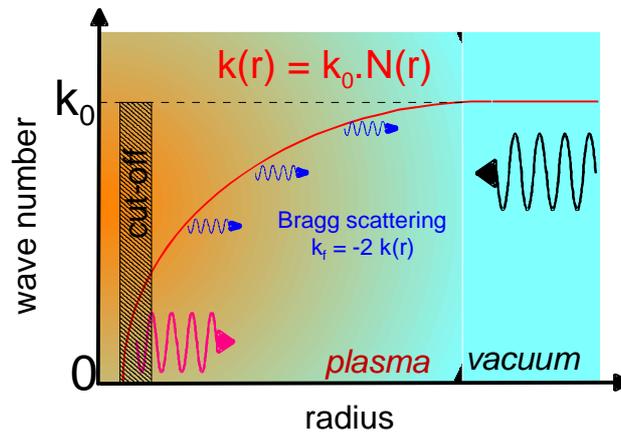

*Figure 1 : Schematic of the propagation wave through plasma. As the wave propagates towards the cut-off layer the refractive index (N) goes to zero. During its propagation the wave experience Bragg scattering processes at high k values, while close to the cut-off the propagation is mainly affected by local density gradients and low k turbulence.*

These fluctuations bring phase and amplitude fluctuations onto the recorded signal. Figure 2 is an example of the variation of the time of flight (ToF) $t = \dfrac{1}{2\pi}\dfrac{\partial f}{\partial F}$, of the propagating probing wave. Thus, the derivative of the phase clearly exhibits the phase fluctuations induced by the density fluctuations.

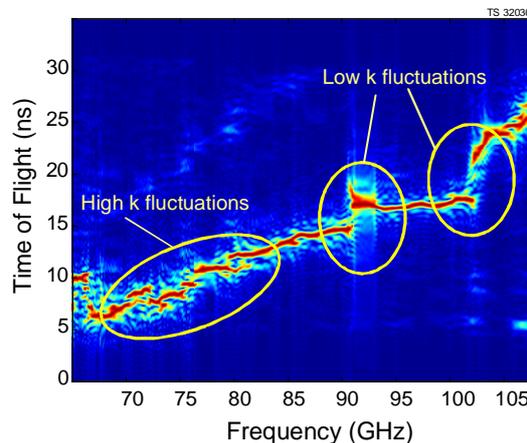

*Figure 2 : Time of flight (phase derivative) of the probing electromagnetic wave propagating through the plasma. Two kinds of perturbations (large and small scale density fluctuations) can be identified.*

Despite the continuous variation of the probing wavenumber value we will classify these fluctuations into two categories, large size (low $k$) and small size (high $k$) turbulence, which will be described in the following sections.

### III. MHD

On tokamaks particles follow a helical path as they go round the torus. This helical path is determined by the safety factor $q(r)$ that is the ratio between the toroidal and the poloidal rotations round the torus. $q(r)$ is itself strongly related to the plasma current profile $q(r) \alpha \frac{B_\Phi(r)}{I(r)}$ ($B_F$ and $I$ being respectively the toroidal field and current). MHD equilibrium of tokamak configuration exhibits singular magnetic surfaces when the safety factor profile reaches rational values. At these positions, magnetic field line reconnections give rise to the formation of magnetic islands. High spatial and temporal resolution of fast sweep reflectometry makes possible to investigate the MHD activity, to localise the islands and their dynamics up to 20 kHz [3]. Furthermore, an analysis of the amplitude variations of the reflected signal points out precious indications on their poloidal shape. There is a special interest in identifying modes with low values of the integers m and n in order to constraint equilibrium code during calculations of the plasma current profile.

Experimental observation and radial localisation of magnetic islands is performed through the detection of the major group delay jumps. Figure 3a and 3b illustrates the $q$ profile evolution recorded by the reflectometry during a plasma current ramp-up experiment. Localised perturbations are radially detected and agree with the $q=2/1$ and $q=1/1$ rational surfaces calculated by the equilibrium code. However, at $t = 9$s, the $q=2/1$ surface is no longer visible and it points out the limit of this method since magnetic islands have to be present into the plasma discharge to be recorded.

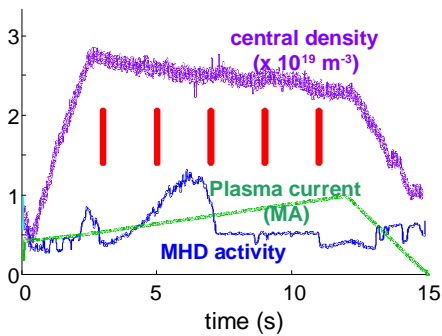

*Figure 3a : Tore Supra plasma discharge where each vertical bars represent 1000 acquisitions of frequency sweeping during a current ramp-up.*

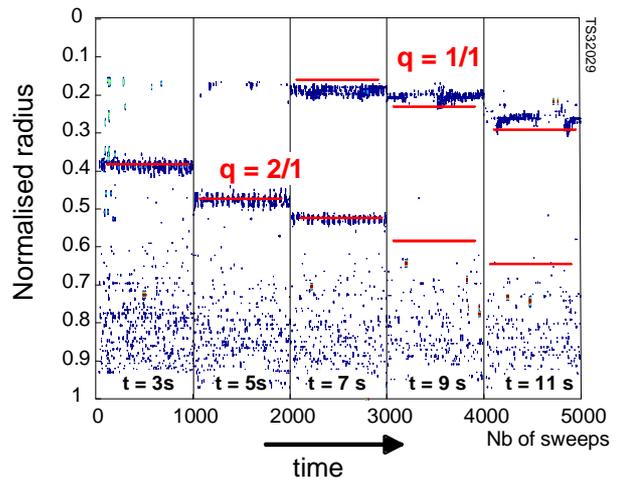

*Figure 3b : Contour plot of detected ToF jumps that coincide with rational q safety factor value radial positions calculated with equilibrium code (horizontal lines).*

## IV. Micro-turbulence

Subtraction of the recorded phase of each frequency sweep from the average phase over several sweeps (typically a few hundreds) provides the reflected signal phase fluctuations (Figure 4a). The radial dependence of the phase fluctuations is determined from the reconstructed density profiles. A Fourier analysis of this radial dependence provides then the wavenumber spectrum of the phase fluctuations (Figure 4b).

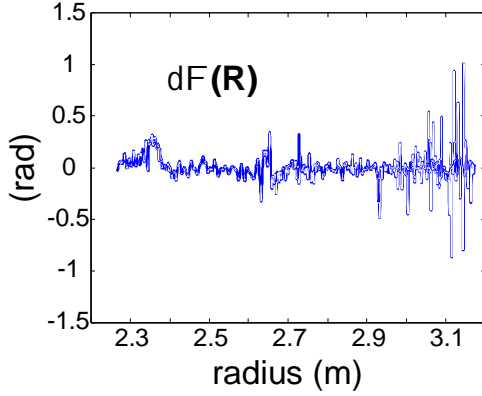 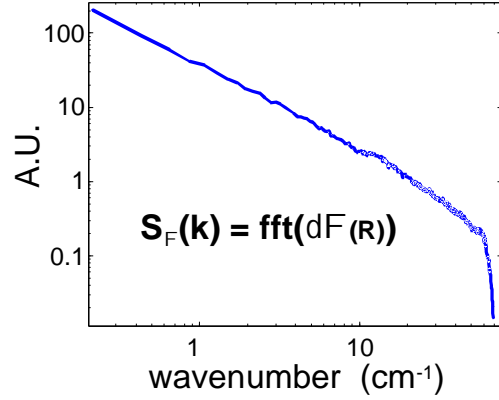

*Figure 4a : radial dependence of the phase fluctuations for one sweep reflected signal.*

*Figure 4b : Wavenumber dependence of the phase fluctuation spectrum.*

Unfortunately, in X-mode, there is no explicit analytical relationship between the phase fluctuations as for O-mode propagation [4] and the electron density fluctuations. Consequently, numerical 1D full wave calculations are requested to provide the normalised transfer function. The inputs of the code are the experimental density profiles, the calculated magnetic field profile and a required model for the density fluctuations $dn(r) = dn_0 \frac{1}{\sqrt{N_k}} \sum_{j=1}^{j=N_k} f(r) g(jk_{min}) \sin(jk_{min} r + f_j)$. This model is the summation of independent modes with given radial (with constant amplitude for the normalisation) and k spectrum distribution (wide enough to cover twice the wavenumber probing wave range).

Moreover, when performing radial sliding Fourier transform (Figure 5a) we can get some radial dependence of the phase fluctuations as long as we consider the spectrum between a restricted wavenumber domain (typically between 3 to 10 cm$^{-1}$). As a matter of fact, Bragg back scattering occurs all along the propagating plasma region of the probing wave, but one can get reasonable spatial resolution (ΔR ~ 3 to 5 cm depending on local gradients) for such fluctuating wavenumber range. Thus, the density fluctuations (Figure 5b) are determined by performing the power spectral density (PSD) of the phase fluctuation spectrum corrected by the transfer function *F(k)* determined with the full wave code.

$$<dn_e^2> = \frac{1}{k_2 - k_1} \int_{k_1}^{k_2} S_{d\Phi}(k) F(k) dk$$

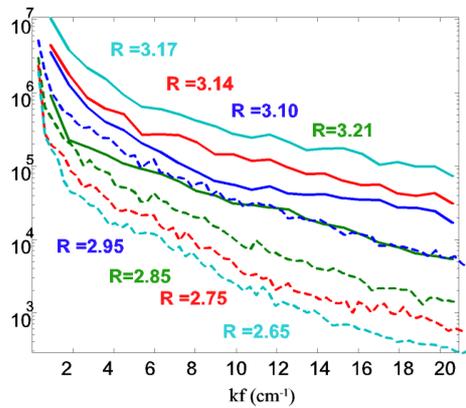 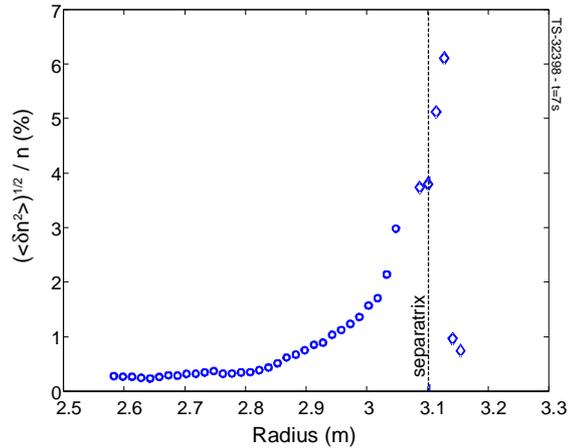

*Figure 5a : Radial evolution of the wavenumber spectrum of the phase fluctuation from sliding Fourier analysis.*

*Figure 5b : Radial profile of the density fluctuation for ohmic plasma discharge from PSD of the sliding phase spectra.*

## V. Summary

Fast frequency sweeping reflectometry has been routinely running to perform density profile measurements onto tokamak devices for many years now. Due to the high technical performances of Tore Supra set-up, accurate determination of the reflected phase signal is achieved with a high radial resolution. MHD activity can be clearly observed as long as magnetic islands are present into the plasma. This situation gives relevant experimental information to constrain the plasma current profile calculated from equilibrium codes. Last but not least, the radial profile of the density fluctuations has been determined. This new and very promising result need further experiment in order to bring additional information onto the energy confinement issues required to burning plasma experiment.